\documentclass[aps,floats,prb,showpacs,twocolumn]{revtex4}

\newcommand{\beq}{\begin{eqnarray}} \newcommand{\eeq}{\end{eqnarray}}

\def\eq{&=&}

\usepackage{amsfonts,amsmath} \usepackage{bm} \usepackage{dcolumn}
\usepackage{epsfig} \usepackage{latexsym}

\begin{document}

\title{Electron transport in two--dimensional arrays}

\author{Julia S. Meyer$^1$, Alex Kamenev$^1$, and Leonid I.
  Glazman$^{1,2}$}

\affiliation{$^1$Department of Physics, and $^2$W.I. Fine Theoretical
  Physics Institute,\\
  University of Minnesota, Minneapolis, MN55455, USA}

\date{\today}


\begin{abstract}
  We study charge transport in a granular array with high inter--grain
  conductances. We show that the system exhibits a
  Berezinskii--Kosterlitz--Thouless crossover from the
  high--temperature conducting state into a low--temperature
  insulating state.  The crossover takes place at a critical
  temperature $T_{\rm BKT}\propto E_c\exp\{-g\}$, where $E_c$ is the
  charging energy of a grain and $g\gg 1$ is the dimensionless
  inter--grain conductance. A uniformly applied gate voltage drives
  the insulator into a conducting charge liquid state followed by an
  insulating lattice--pinned Wigner crystal state at larger values of
  the gate voltage. Technically, we establish correspondence between
  the charge and phase representations, employing the instanton gas
  summation in the framework of the phase model.
\end{abstract}

\maketitle

\section{Introduction}

Granular arrays have recently attracted much attention as analytically
tractable systems to study the interplay of interactions and
scattering~\cite{Gerber97,general,Beloborodov01,tschersich,AGK}.  The
advantage of granular systems is the possibility to separate the
scattering--induced quantum interference phenomena from
electron--electron interaction effects. Quantum coherence is relevant
as long as the typical dwell--time in a grain $\tau_{\rm dwell}\sim
\hbar/(g\delta)$ is longer than the dephasing time $\tau_\varphi$.
Here $g$ is the dimensionless inter--grain conductance measured in
units of $e^2/(2\pi\hbar)$, and $\delta$ is the mean level spacing of
the grains.  For two--dimensional disordered interacting systems the
dephasing time is known to be~\cite{AAK82} $\tau_\varphi\approx \hbar
g/T$.  The quantum interference is thus suppressed if $\tau_\varphi <
\tau_{\rm dwell}$, which is the case at $T>g^2\delta$.  Hereafter we
assume this condition to be satisfied and consider the temperature
range $g^2\delta<T<E_c$ (where $E_c=e^2/(2C)$ is the charging energy
of the grains). This allows us to focus on the interaction--induced
phenomena, while omitting the interference (incoherent regime).

Earlier studies of incoherent two--dimensional arrays led to
conflicting theoretical results~\cite{tschersich,fs2,fs}. In
Refs.~[\onlinecite{fs2,fs}] the low--temperature insulating state was
found for sufficiently small inter--grain conductance, $g<1$. Upon
elevating the temperature, the array was shown to undergo a
Berezinskii--Kosterlitz--Thouless (BKT) transition into the conducting
state. The transition temperature $T_{\rm BKT}=T_{\rm BKT}(g)$ was
predicted to go to zero at some critical value of conductance
$g_c\approx 1.8$. At larger conductances, $g> g_c$, the metallic state
was claimed to persist down to zero temperature. In
Ref.~[\onlinecite{tschersich}] the same model as in
Refs.~[\onlinecite{fs2,fs}] was studied in the regime of large
conductances, $g\gg 1$. Using a perturbative renormalization group
(RG) analysis, the renormalized inter--grain conductance was shown to
behave as $g\Rightarrow g(T)=g-{2\over d}\ln(E_cg/T)$ (where $2d$ is
the coordination number of the lattice).  This correction is
essentially similar (and at $T\approx g\delta$ crosses
over~\cite{vinokur}) to the interaction--induced
Altshuler--Aronov~\cite{altshuler-aronov} conductivity corrections
known for homogeneous disordered systems. At $T\sim E_c
g\exp\{-dg/2\}$, the conductance is renormalized down to $g(T)\sim
{\cal O}(1)$. Thus, one may expect that the system approaches an
insulating state at low enough temperature even for large bare
inter--grain conductance.  Whether such a ``high--$g$'' insulator
indeed exists, and -- if so -- its nature and relation (if any) to the
BKT transition, found~\cite{fs2,fs} for ``low--$g$'' systems, was not
clarified.

In the present work we show how to reconcile these findings. In
particular, we show that there exists a finite $T_{\rm BKT}\propto
\exp\{-g\}$ at $g\gg1$. To this end one needs to go beyond the
perturbative analysis~\cite{tschersich} of the model considered in
Refs.~[\onlinecite{tschersich,fs2,fs}] and include non--perturbative
-- instanton -- field configurations. A similar program was recently
carried out for one--dimensional incoherent arrays~\cite{AGK}. The
conductivity of one--dimensional arrays was found to display activated
(insulating) behavior with the charge gap $\sim E_c\exp\{-g/4\}$,
which is parametrically larger than the temperature where the
perturbative corrections become large.  It was also shown that the
proper low--temperature representation of incoherent arrays is that of
pinned {\em charge}--density wave fluctuations.  [This should be
contrasted with the fluctuating phase (or voltage) picture employed in
Refs.~[\onlinecite{tschersich,fs2,fs}].]  The activation gap
corresponds to the energy needed to create a long unit--charge
soliton.

In the present paper the charge representation is derived and analyzed
for the two--dimensional setup. We find that the charge excitations
are localized unit--charge two--dimensional solitons. At $g\gg 1$ the
solitons interact logarithmically over a large range of distances.
This leads to a sharp BKT crossover~\cite{BKT-b,BKT-kt} between a
low--temperature insulating phase with bound charge--anti-charge pairs
(and an exponentially small number of free charges) and a
high--temperature conducting phase, where the pairs are unbound. The
BKT temperature $T_{\rm BKT}(g)$ remains finite (though exponentially
small), $T_{\rm BKT}(g)\sim E_c g \exp\{-g\}$, for an {\em arbitrarily
  high} bare conductance $g$. The zero--temperature quantum phase
transition at $g=g_c$, found in Refs.~[\onlinecite{fs2,fs}], thus,
does {\em not} exist. Instead, there is a fast but continuous drop of
the transition temperature $T_{\rm BKT}(g)$ in the vicinity of $g\sim
1$.

The issue of whether a classical phase transition or a crossover
occurs as the temperature is lowered, depends sensitively on the
details of the model. The true BKT transition takes place only if the
interactions between the charged solitons are logarithmic at
arbitrarily large distances. This is the case for arrays with
inter--grain capacitances only (in the absence of the grain's
self--capacitance, no electric field lines can leave the 2d plane).
In the presence of the self--capacitance, the interaction is
logarithmic in a wide, but finite range of distances:
$1<l\lesssim\exp\{g/2\}$ (hereafter distances are measured in units of
the array lattice spacing). In the latter case, below the BKT
temperature the array's conductivity is not zero (in contrast to the
former case), but rather exhibits the activation behavior,
\begin{equation}
                                                      \label{eq-res}
\sigma\simeq g\exp\left\{-\frac{\Delta}T\;\right\},
\end{equation}
where the activation gap is given by $\Delta \simeq g T_{\rm BKT}\gg
T_{\rm BKT}$. Upon raising the temperature, above the BKT temperature
the conductivity sharply increases as
\begin{equation}
                                                      \label{eq-above}
\sigma=gK\exp\left\{-2b\sqrt{\frac{T_{\rm BKT}}{T-T_{\rm BKT}}}\right\},
\end{equation}
where $K$ and $b$ are non-universal constants of order
unity~\cite{mooij}.  Finally, above the transition region, the
conductivity crosses over to the perturbative prediction
$\sigma=g-\ln(gE_c/T)$. We thus have a generally consistent picture
based on the BKT physics at any value of bare conductance, $g$.

A gate voltage induces a uniform background charge $q\in [0,1]$.
However, for small gate voltages the array remains in the
particle-hole symmetric state with an integer number of electrons per
dot. The transition into a non--uniform state (with a non--integer
average number of electrons per dot) takes place at a critical
dimensionless charge density $q^*=\Delta/(2E_c)$.  Its physics is
similar to the transition from the Meissner to the Abrikosov state in
type II superconductors upon increasing an external magnetic field. In
our case the role of the magnetic field is played by the gate voltage,
$q$, with $q^*$ corresponding to the lower critical field $H_{c1}$.
The Abrikosov lattice in turn corresponds to the 2d Wigner crystal
formed by the unit--charge solitons. Such a crystal is easily pinned
by the lattice.  Consequently, at low temperatures, the array is in
the insulating phase with a residual activation conductivity,
associated with the thermal creation of defects.  The Wigner crystal
melts at a temperature~\cite{Tm} of the order of $T_{\rm BKT}$,
leading to a sharp crossover into the conducting phase.

Methodologically, interacting systems may be modeled in two
alternative ways: in terms of either phase or charge degrees of
freedom. The two are canonically conjugated and the choice between
them is a matter of convenience. The phase representation is easier to
derive microscopically starting from the fermionic tunneling
Hamiltonian. By this reason it was used in the vast majority of works
on granular systems and quantum
dots~\cite{general,Beloborodov01,tschersich,fs,fs2}.  We found it more
convenient, however, to work in the charge representation, which is
the natural language to describe the insulating phase. In the quantum
dot context, the charge description was introduced in
Refs.~[\onlinecite{flensberg,matveev}]. Here we employ its
generalization to 2d granular arrays. We introduce the model in
Sec.~\ref{sec-charge} and show that it exhibits the BKT crossover in
Sec.~\ref{sec-BKT}. Finite gate voltages and the pinned Wigner crystal
phase are discussed in Sec.~\ref{sec-gate}. To facilitate comparison
with the body of work on the phase representation, we include the
proof of equivalence of the two models in Appendix~\ref{app-phase}.

\section{Charge representation}
\label{sec-charge}

In this section, we introduce the charge representation for incoherent
($\delta\to 0$) interacting arrays. To keep the presentation compact,
we assume all contacts to be single--channel, characterized by a
reflection amplitude $r<1$. The generalization to the multichannel
case and, in particular, to $g\gg1$ is discussed at the end of the
section and, in more detail, in Appendix~\ref{app-gamma}, while the
proof of the equivalence of the resulting charge model to the more
widely used phase model is outlined in Appendix~\ref{app-phase}.

\subsection{Single contact}

Consider a point contact between a quantum dot and a metallic
reservoir. Such a point contact allows for a small number of
propagating transverse modes which may be thought of as
one--dimensional electron liquids (with the contact situated at the
origin, $z=0$). Here we consider the case of a single propagating
mode, deferring the consideration of multi--mode contacts to
Appendix~\ref{app-gamma}. The corresponding one--dimensional electron
liquid may be bosonized in the conventional
way~\cite{flensberg,matveev} and described in terms of the bosonic
field $\theta(\tau,z)$. Its gradient $\partial_z \theta(\tau,z)$ has
the meaning of a local electron density. As a result, the electron
number on the dot may be written as $N=\int_0^\infty\! dz\, \partial_z
\theta(\tau,z)=-\theta(\tau,0)$ and, thus, the Coulomb energy takes
the form $(eN)^2/(2C)=E_c\theta^2(\tau,0)$. Finally the
imaginary--time action of the bosonic field reads
\begin{eqnarray}
                                                    \label{eq-dot}
S[\theta(\tau,z)] &= &\int\limits_0^\beta d\tau
  \left\{ \int\limits_{-\infty}^{\infty}dz
  \left[(\partial_\tau\theta)^2+(\partial_z\theta)^2\right]\right.
   \nonumber \\
  && + \left. E_c\theta^2(\tau,0) -
\frac{Dr}\pi\cos[2\pi\theta(\tau,0)]\right\}.
\end{eqnarray}
The last term in this expression describes backscattering at the point
contact with the reflection amplitude $r$, while $D$ is the electronic
bandwidth.

One may integrate out all degrees of freedom with $z\neq 0$, retaining
the field $\theta(\tau)\equiv \theta(\tau,0)$ only. The corresponding
action reduces to
\begin{equation}
                                               \label{eq-dot1}
S[\theta] = \frac1T\sum_m \left( \pi|\omega_m|\theta_{m}^{2}+
E_{\rm c}\theta_{m}^2 \right) - \frac{Dr}\pi\!\int\limits_0^\beta
\!d\tau\,\cos(2\pi\theta(\tau)),
\end{equation}
where $\omega_m=2\pi Tm$, and we have introduced the Matsubara
representation though the transformation $\theta_m=\int_0^\beta
d\tau\, \theta(\tau)e^{-i\omega_m\tau}$. The dissipative term,
$\pi|\omega_m|\theta_{m}^{2}$, is generated as a result of integrating
out the continuum spectrum of the degrees of freedom on the dot.  Its
appearance is a consequence of the assumption that the mean level
spacing is the smallest energy scale in the model, $\delta\to 0$.

\subsection{2d array}
\label{subsec-array}

\begin{figure}[h]
  \centerline{\epsfxsize=3in\epsfbox{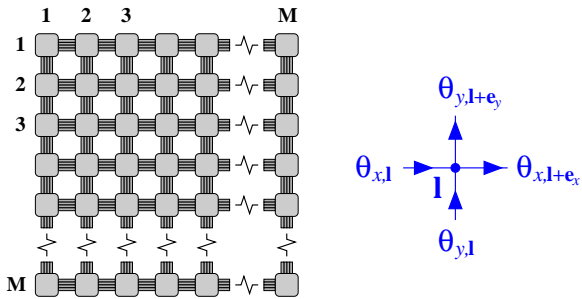}}
 \caption{Granular array with a square lattice. The
   massless modes $\eta$, explained in the text, correspond to
   circular currents around a plaquette and, therefore, do not
   contribute to charge transport.}
\label{fig1}
\end{figure}

We now generalize the single--contact action, Eq.~(\ref{eq-dot1}), to
the 2d array geometry. To this end we introduce the vector index ${\bf
  l}$ to label the grains. We also introduce two fields
$\theta_{x,{\bf l}}(\tau)$ and $\theta_{y,{\bf l}}(\tau)$ which
describe charge transport from grain ${\bf l}$ in the positive $x$ and
$y$ directions, respectively. In these notations, the instantaneous
electron density on the grain ${\bf l}$ is given by the lattice
divergence $\nabla\cdot\vec\theta_{\bf l} \equiv \theta_{x,{\bf
    l+e}_x}-\theta_{x,{\bf l}}+\theta_{y,{\bf l+e}_y}-\theta_{y,{\bf
    l}}$ (cf.~Fig.~\ref{fig1}). With the backscattering in the
contact, characterized by the reflection amplitude $r$, the action
reads
\begin{widetext}
\begin{equation}
\label{eq-matveev} S\!\left[\,\vec\theta\,\right] = \sum_{\bf l}
\left\{\frac1T\sum_m \left( \pi|\omega_m|\vec\theta_{{\bf
l},m}^{\;2} + E_{\rm c}(\nabla\cdot\vec\theta_{{\bf l},m})^2
\right) - \frac{Dr}\pi\sum_{i=x,y}\int\limits_0^\beta d\tau\;
\cos(2\pi\theta_{i,{\bf l}}(\tau))\right\},
\end{equation}
\end{widetext}
where $D$ is again the bandwidth. As in Eq.~(\ref{eq-dot1}), the first
term in the action~(\ref{eq-matveev}) describes the dissipative
dynamics originating from integrating out degrees of freedom within
the grains, the second term is responsible for the charging, and the
third one describes backscattering in the contacts.

For a square lattice of linear size $M$, the array contains $M^2$
grains and $2M^2$ contacts between them. Consequently the model is
written in terms of $2M^2$ bosonic degrees of freedom, $\theta_{i,{\bf
    l}}(\tau)$.  In the limit $r\to 0$, the masses of these modes are
provided only by the $M^2$ charging terms $E_{\rm
  c}(\nabla\cdot\vec\theta_{{\bf l},m})^2$. Therefore, if the
backscattering is neglected, only half of the degrees of freedom of
the model are massive. To see this explicitly, one may rewrite the 2d
vector field $\vec \theta$ through two scalar fields
$\vec\theta=\nabla{\chi}+\nabla\times\eta$ and notice that the
charging term contains only the field $\chi$ while the curl--field
$\eta$ fully decouples from it.

In order to find an effective low-energy theory, we shall proceed with
the renormalization group (RG) scheme based on the integration of the
high frequency Matsubara modes, $D'<|\omega_m|<D$, accompanied by the
appropriate change of the backscattering amplitude $r$. As long as the
coefficient in front of the cosine--term in Eq.~(\ref{eq-matveev}) is
less than the running bandwidth $D'$, one may treat the fields as
Gaussian, governed by the first two terms in Eq.~(\ref{eq-matveev}).
As a result, the backscattering amplitude renormalizes as
\begin{equation}
                                         \label{eq-renormalization}
Dr\enspace\Rightarrow\enspace Dr \exp\left\{-{(2\pi)^2\over 2}\langle
\theta^{2}_i\rangle\right\}\, ,
\end{equation}
where the averaging in $\langle \theta^{2}_i\rangle$ is performed over
high--frequency fluctuations. Passing to the momentum representation
and taking into account both $\chi$ and $\eta$ components of the
fluctuations, one finds
\begin{eqnarray}
\langle\theta^{2}_i\rangle\! &=& \!\frac
T{4M^2}\!\!\!\sum\limits_{|\omega_m|=D'}^{D}\sum\limits_{
q_x,q_y=1}^{M}\left(\frac1{E_{\bf
      q}\!+\!\pi|\omega_m|}\!+\!\frac1{\pi|\omega_m|}\right)\!
\nonumber \\
       &\simeq&
\!\frac1{4\pi^2}\left(\ln\frac{ D}{E_c}+\ln\frac{D}{D'}
\right)=\frac1{2\pi^2}\ln\frac D{\sqrt{E_c D'}}\, ,
                                        \label{eq-theta2}
\end{eqnarray}
where $E_{\bf q}=4E_c\sum_{i=x,y}\sin^2(\pi q_i/(2M))$ is the mass
spectrum of the $\chi$ modes. In the second line we have assumed that
$D'<E_c$ (in the opposite case, $D'>E_c$, one should substitute
$\sqrt{E_c D'}$ by $D'$). Notice that the presence of the lower limit,
$D'$, in this expression is due to the massless rotational modes of
the field $\eta$. [Note also the difference with the 1d--system, where
all modes are massive and, therefore, the result corresponding to
Eq.~(\ref{eq-theta2}) is independent on the lower limit~\cite{AGK}.]
Combining Eqs.~(\ref{eq-renormalization}) and (\ref{eq-theta2}), one
finds that upon integrating out the high--frequency modes, the
coefficient of the cosine potential renormalizes as $Dr \Rightarrow
\sqrt{E_cD'}\,r$. As was discussed above, this procedure works as long
as $\sqrt{E_cD'}\,r<D'$, that is $D'>T_0$, where $T_0= E_c r^2$ is the
``freezing'' temperature. For smaller bandwidths, the cosine--term
itself provides a mass for the rotational modes $\eta$. As a result,
all modes acquire a mass and, thus, the renormalized backscattering
amplitude looses its sensitivity to the lower limit $D'$. Therefore,
we arrive at the conclusion that for $D'< T_0$ the cosine amplitude
saturates at a value about $T_0$.  Integrating in this way all
Matsubara components, except the static one, $m=0$ (it is obvious from
Eq.~(\ref{eq-theta2}) that the $m=0$ component can not be handled in
the same way), one obtains an effective classical model with the
action
\begin{equation}
                                                   \label{eq-classical}
S_{\rm cl}[\vec\theta] = \frac{E_c}T \sum_{\bf l}
\Big\{(\nabla\cdot\vec\theta_{\bf l})^2 -
\frac{\gamma(T)}{2\pi^2}\sum_{i=x,y}\cos(2\pi\theta_{i,{\bf
l}})\Big\},
\end{equation}
where $\gamma(T)=2\pi \sqrt{T/E_c}\,\,r$ for $T>T_0$ and
$\gamma(T)\simeq 2\pi r^2$ for $T<T_0$.

So far we have formulated the model for an array with single--channel
contacts. Generalization to the $N\geq 2$ channels is achieved by
introducing bosonic modes $\vec\theta_{{\bf l},\alpha}(\tau)$ for
every channel $\alpha=1\dots N$. One may then integrate out the
antisymmetric modes (including the spin modes) for every contact,
retaining only the symmetric (charge) mode $\vec\theta_{{\bf
    l}}=\sum_{\alpha=1}^N \vec\theta_{{\bf l},\alpha}$, see
Appendix~\ref{app-gamma} for details. The main result of such a
procedure may be summarized by the redefinition of the effective
backscattering amplitude~\cite{nazarov,ABG} $r\Rightarrow c_N
\prod_{\alpha=1}^N r_\alpha$ in the action of the charge mode ($c_N$
is a numerical coefficient with a finite limit $c_\infty$).
Consequently the characteristic freezing temperature changes to
$T_0\simeq E_c \prod_{\alpha=1}^N r^2_\alpha$. Then, the charge mode
may be described by the same effective classical model
Eq.~(\ref{eq-classical}) with $\gamma(T)\simeq \sqrt{T/E_c}\,
\prod_{\alpha=1}^N r_\alpha$ for $T>T_0$ and $\gamma(T)\simeq
\prod_{\alpha=1}^N r^2_\alpha$ for $T<T_0$.

A model which adequately describes an array of metallic grains assumes
that the contacts between grains consist of a large number $N$ of
weakly transmitting
channels~\cite{general,Beloborodov01,tschersich,fs,fs2}. For
sufficiently large $N$, the total conductances of the junctions may
still be high, $g=\sum_{\alpha=1}^N t_\alpha^2 >1$, where
$t_\alpha^2=1-r_\alpha^2\ll 1$ is the transmission probability in
channel $\alpha$. In this case, one finds that $\prod_\alpha
r_\alpha=\exp\left\{\frac12\sum_\alpha\ln(1-t_\alpha^2)\right\}\approx
\exp\{-g/2\}$. Employing the expressions derived above, one obtains
that the freezing temperature is of the order $T_0\simeq g E_ce^{-g}$,
while the effective amplitude of the cosine potential in
Eq.~(\ref{eq-classical}) is given by
\begin{eqnarray}
                                               \label{eq-gamma}
 \gamma(T)\simeq
\begin{cases}
  \sqrt g\,\, e^{-g/2}\sqrt{\frac T{E_c}}\,; \,\,\,\, & T>T_0,\\
  g \,e^{-g}\,; & T<T_0\, .
\end{cases}
\end{eqnarray}
In Appendix~\ref{app-phase}, we show how Eqs.~(\ref{eq-classical}) and
(\ref{eq-gamma}) follow from the phase
model~\cite{general,Beloborodov01,tschersich,fs,fs2}, demonstrating
that the two models based on the charge and phase representations,
respectively, are reduced to the same effective classical system.
Notice that, since the charge model was derived for $r_\alpha\ll 1$,
while $t_\alpha\ll 1$ is assumed in the phase model, the coincidence
of $\gamma(T)$ may be expected at best with exponential accuracy. The
algebraic pre--exponential function of $g$ is a result of the
evaluation in the framework of the phase model, see
Appendix~\ref{app-phase}.

\section{BKT transition}
\label{sec-BKT}

In this Section we analyze the physics of the classical charge model
specified by Eqs.~(\ref{eq-classical}) and (\ref{eq-gamma}).  Two
issues are discussed: (i) the spectrum of its charged excitations and
their interactions, and (ii) the low--frequency charge dynamics and
the dc conductivity of the array. We finally put our findings in
perspective by comparing them with the results of previous studies.

\subsection{Charge spectrum}

The lowest energy configuration of the action (\ref{eq-classical}) is
given by $\vec\theta=0\;(\mbox{mod}\;1)$ everywhere.  Localized
excitations must have integer $\vec\theta$ far away from the core to
minimize the cosine potential. The total charge of such localized
excitation is $e\!\int(d^2l)\,\nabla\cdot\vec\theta=e\!\int d\vec
s\cdot\vec \theta$, where the line integral on the r.h.s. is
calculated over a distant contour enclosing the excitation. It is
clear therefore that the charge of the excitation is quantized in
integer numbers of $e$. The simplest (and only stable) charged
excitations have charge $\pm e$.  They consist of a large (i.e.,
spread out over $\sim 1/\gamma\gg 1$ grains) localized 2d soliton of
unit charge, connected to a 1d string of links with $\theta_i=1$.  The
other end of the string may either go to the system boundary or be
terminated by an anti-soliton with charge $-e$. The soliton solution
centered at ${\bf l}=0$ can be written in the form $\vec\theta_{\bf
  l}=1-\vec\vartheta({\bf l})$ for the links along the string and
$\vec\theta_{\bf l}=\vec\vartheta({\bf l})$ everywhere else, where
$|\vec\vartheta(|{\bf l}|\to\infty)|\to0$.  Minimizing the action,
Eq.~(\ref{eq-classical}), with respect to $\vec\vartheta$, one finds
the saddle point equation for the soliton solution,
\begin{equation}
\nabla(\nabla\cdot\vec\vartheta)-{\gamma\over 2\pi} \sum_{i=x,y}\sin
(2\pi\vartheta_i){\bf e}_i=0.
\label{saddlepoint}
\end{equation}
Except for a domain consisting of ${\cal O}(1)$ links closest to the
core of the soliton, $\vartheta$ is small, justifying an expansion of
the sine--term in the saddle point equation. As a result,
Eq.~(\ref{saddlepoint}) takes the form
$\nabla(\nabla\cdot\vec\vartheta)-\gamma\vec\vartheta=0$. Its
unit--charge solution is:
\begin{eqnarray}
                                     \label{eq-soliton}
\vec\vartheta({\bf l}) \eq -\frac{\sqrt\gamma}{2\pi}\,
K_1\!\left(\frac l{\xi_{\rm
  s}}\right)\,{\bf e}_l,
\end{eqnarray}
where $K_1$ is a modified Bessel function, and $\xi_{\rm
  s}=1/\sqrt\gamma \gg1$, justifying the continuum approximation;
finally ${\bf e}_l\equiv {\bf l}/l$. The solution is normalized as
$\int(d^2l)\,\nabla\cdot\vec\theta=1$ to obey the charge quantization.

Substituting this solution back into the action,
Eq.~(\ref{eq-classical}), one finds that the soliton energy originates
primarily from the cosine potential part of the action and is given by
$\Delta\simeq E_c[\gamma(T)/2\pi]\ln \xi_{\rm s}$. The large
logarithmic factor $\ln \xi_{\rm s}=-{1\over 2} \ln\gamma\gg 1$ is due
to the $\propto 1/l^2$ behavior of the charge density in the wide
range of distances $1<l<\xi_{\rm s}$. At larger distances, $l>\xi_{\rm
  s}$, the charge density decays exponentially. As a result, the
solitons interact logarithmically up to a distance $\xi_{\rm s}$
beyond which the interaction is exponentially screened.  Since the
density of thermally--excited solitons is $n_{\rm s}\approx
\exp\{-\Delta/T\}$, the mean distance between them is $l_{\rm
  s}=n_{\rm s}^{-1/2}\approx \exp\{\Delta/(2T)\}$. It becomes
comparable to $\xi_{\rm s}$ at $T\approx \Delta/(2\ln \xi_{\rm s})=
E_c\gamma(T)/(4\pi)$.  This condition is satisfied at temperatures
about the ``freezing'' temperature, $T\sim T_0$.  Thus, at $T < T_0$,
the thermally-excited charges are essentially non--interacting, while,
at $T>T_0$, there is a neutral (in average) gas of logarithmically
interacting solitons and anti-solitons.

In the latter regime the partition function of the charged degrees of
freedom may be written therefore as
\begin{equation}
                                       \label{eq-partition}
Z=\sum\limits_{n=0}^{\infty}\frac{f^{n}}{n!} \int (d^2l_1)\ldots
(d^2l_n)\,e^{\pm\frac{E_c\gamma(T)}{2\pi T}\sum\limits_{k,k'}^n
\ln|{\bf l}_k-{\bf l}_{k'}|}\, ,
\end{equation}
where $\ln f\simeq E_c\gamma(T)/T$ is the fugacity of the logarithmic
gas, originating from the solitons core energy. The plus/minus signs
in the exponent correspond to soliton--soliton and
soliton--anti-soliton interactions, respectively.

It is well known that the Coulomb gas in 2d described by
Eq.~(\ref{eq-partition}) undergoes the BKT
transition~\cite{BKT-b,BKT-kt} at a critical temperature $T_{\rm
  BKT}\approx E_c\gamma/(4\pi)$.  For $T<T_{\rm BKT}$, the charges are
bound in charge--anti-charge pairs. The residual density of free
charges is exponentially small and given by $n_{\rm s}\approx
\exp\{-\Delta/T\}$, where $\Delta=T_{\rm BKT}\ln \xi_{\rm s}^{\;2}$.
The value of $\Delta$ is finite but large, as long as the solitons
interact with each other logarithmically over a broad range of
distances $\xi_{\rm s}\gg 1$. Notice that the Coulomb interactions in
our model are strictly on--site (only the self--capacitance, $C$, is
included). The long range of the soliton--soliton interactions is due
to the fact that in a strongly coupled array, $g\gg 1$, the charge is
spread over a large distance $\xi_{\rm s}\sim\exp\{g/2\}$.

One can modify the model to include mutual capacitances $C'$ between
neighboring grains (and thus to include long--range Coulomb
interactions). It is straightforward to show that such modification
alters the range of logarithmic interactions as $\xi_{\rm s} \to
\xi_{\rm s}\sqrt{1+C'/C}$, while the charging energy now reads
$E_c=e^2/(2(C+C'))$.  In the limit $C\to 0$, while $C'$ remains
finite, the interaction range diverges, $\xi_{\rm s}\to \infty$. In
fact, this was to be anticipated: since without the self--capacitance
no electric field lines can leave the system, one deals with the true
2d Coulomb interaction, which is logarithmic. In this case $\Delta\to
\infty$ and the density of free charges below $T_{\rm BKT}$ is
strictly zero. This is the case of the genuine BKT phase transition.
For non--vanishing self--capacitance, $C>0$, the interactions are
screened at distances exceeding $\xi_{\rm s}$.  Therefore, the density
of free charges is finite at any temperature and the phase transition
is smeared into a sharp crossover. Above the transition/crossover
temperature the density of free charges rapidly increases
as~\cite{BKT-kt} $n_{\rm s}\sim\exp\{-2b\sqrt{T_{\rm BKT}/(T-T_{\rm
    BKT})}\}$, where $b$ is a constant of order unity, driving the
array into the conducting phase.

\subsection{dc conductivity}

In order to discuss the dc conductivity of the array, one needs to
restore the low frequency, $\omega\ll T$, dynamics of the classical
charge model, Eq.~(\ref{eq-classical}). This may be done formally by
keeping the dissipative dynamical term in the action. Notice that in
the multichannel case, see Appendix~\ref{app-gamma}, the coefficient
in front of $|\omega_m|$ acquires a factor $N^{-1}$, where $N$ is the
number of channels. In the presence of strong backscattering, it
actually reads $\pi g^{-1} |\omega_m|\vec\theta_{{\bf l},m}^{\;2}$ and
corresponds to the conventional Ohmic dissipation.  Since we focus on
the low frequencies, it is convenient to pass to the Keldysh
representation (to avoid dealing with the analytical continuation) and
consider its semi--classical limit. The latter is known to be
equivalent to a certain Langevin dynamics~\cite{Kamenev01}.

Here we prefer to take a more phenomenological route, leading to the
same conclusions. Let us consider the static equations of motion
following from Eq.~(\ref{eq-classical}): $\partial_i\left({e\over
    C}\nabla\cdot\vec\theta_{\bf l}\right)-{e\gamma\over 2\pi C}\sin
(2\pi\theta_{i,{\bf l}})=0$. Since $\frac eC\nabla\cdot\vec\theta_{\bf
  l}\equiv V_{\bf l}$ is the voltage on grain ${\bf l}$, the equation
simply expresses the fact that in the absence of charge quantization,
$\gamma\to 0$, all grains are equipotential: $\nabla V_{\bf l}=V_{{\bf
    l+e}_i}-V_{\bf l}=0$. Once currents are allowed to flow in the
array this condition should be substituted by the Kirchhoff law,
$V_{{\bf l+e}_i}-V_{\bf l}=RI_{i,{\bf l}}$, where $R=2\pi \hbar /(e^2
g)$ is the contact resistance, and $I_{i,{\bf l}}=e\partial_t
\theta_{i,{\bf l}}$ is the current flowing between grains ${\bf l}$
and ${\bf l+e}_i$. Restoring also the $\gamma$--term in the equation
of motion, one thus finds
\begin{eqnarray}
                                                \label{eq-langevin}
&&\frac{\pi}{g}\, \partial_t\vec\theta -
E_c\Big[\nabla(\nabla\!\cdot\!\vec\theta) - \frac{\gamma}{2\pi}
\sum_i {\bf e}_i\sin(2\pi\theta_i)\Big] \nonumber\\
 &=& -\frac{e}{2}\, {\bf E} +\vec\xi(t)\,.
\end{eqnarray}
On the right hand side we have included an external electric field
${\bf E}$, as well as the Gaussian noise, $\vec\xi(t)$, with the
correlator
\begin{equation}
                                            \label{eq-noise}
\langle \xi_{i,{\bf l}}(t)\xi_{i',{\bf l'}}(t')\rangle  = \frac{2\pi
  T}{g}\, \delta(t-t')\delta_{{\bf l},{\bf l}'}\delta_{i,i'}\, ,
\end{equation}
to satisfy the fluctuation--dissipation theorem.

Our goal is to calculate the current, $I$, in presence of a weak
uniform field, $E$. To this end we employ Drude--type arguments,
saying that $I=en_{\rm s} v$, where $n_{\rm s}$ is the carrier
concentration and $v$ is their drift velocity. The only mobile
carriers in the system are the solitons, Eq.~(\ref{eq-soliton}), whose
concentration, $n_{\rm s}$, we have discussed in detail above. Now we
concentrate on the drift velocity, $v$. We look for a solution of
Eq.~(\ref{eq-langevin}) (without the noise) in the form
$\vec\theta({\bf l},t)=\vec\theta_0({\bf l}\!-\!{\bf
  v}t)+\vec\theta_1({\bf l}\!-\!{\bf v}t)+\vec\alpha$. Here
$\vec\theta_0({\bf l})$ is the static soliton solution in the absence
of the external field, whereas $\vec\theta_1\sim {\bf E}$ is a small
modification of the soliton's shape due to the presence of the
external field.  Finally the constant vector $\vec\alpha$ is
determined by the shift of the minimum of the periodic potential in
the field: $E_c\gamma\sum_i{\bf e}_i\sin(2\pi\alpha_i)=\pi{\bf E}$.
Choosing ${\bf E}=E{\bf e}_x$ and ${\bf v}=v{\bf e}_x$, and
linearizing Eq.~(\ref{eq-langevin}), one finds that $\vec\theta_1$
satisfies the equation
\begin{equation}
                               \label{eq-linearized}
E_c \hat {\cal F}_{\{\vec\theta_0\}}\vec\theta_1=\frac
  v{g}\partial_x\vec\theta_0-\frac
  E{\pi}\sin^2(\pi\theta_{0,x}){\bf e}_x,
\end{equation}
where $\hat {\cal F}_{\{\vec\theta_0\}}\vec\theta_1 \equiv
\nabla(\nabla\!\cdot\!\vec\theta_1)-\gamma \sum_i {\bf
  e}_i\cos(2\pi\theta_{0,i})\theta_{1,i}$.  The velocity, $v$, is
determined by the condition that the r.h.s. of
Eq.~(\ref{eq-linearized}) is orthogonal to the translational
zero--mode of the operator $\hat {\cal F}_{\{\vec\theta_0\}}$, given
by $\partial_x\vec\theta_0$. This requirement leads to $v\sim gE$.
Finally, the dc conductivity is given by $\sigma \simeq g\,n_{\rm
  s}(T)$.

As a result, all the conclusions, drawn above, regarding the BKT
transition/crossover in the soliton density may be directly translated
to the array's conductivity. In particular, for the self--capacitance
model at $T<T_{\rm BKT}$ (employing that at low temperature $\ln\gamma
\simeq g$) we find Eq.~(\ref{eq-res}), i.e. $\sigma\simeq
g\exp\{-\Delta/T\}$.  Above $T_{\rm BKT}$, the conductivity behaves as
$\sigma \simeq g\exp\{-2b\sqrt{T_{\rm BKT}/(T-T_{\rm BKT})}\}$, see
Eq.~(\ref{eq-above}).  At even higher temperatures, this behavior
crosses over to the result~\cite{tschersich} of the perturbative
calculation, $\sigma=g-\ln(gE_c/T)$.

\subsection{Phase diagram}

\begin{figure}[h]
  \centerline{\epsfxsize=3.5in\epsfbox{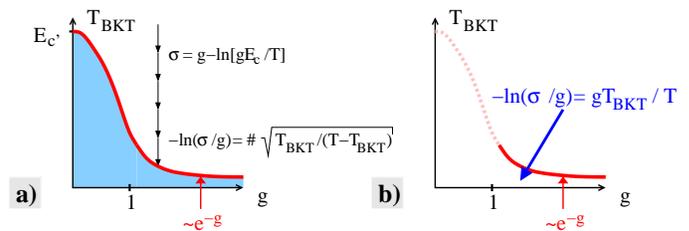}}
 \caption{BKT temperature as a function of $g$. For $g\gg1$, $T_{\rm
     BKT}$ is exponentially small, but remains finite. a) $C=0$: the
   true transition exists for any value of $g$.  Conductance is zero
   below the transition temperature. b) $C>0$: crossover takes places.
   It is sharp only if $g\gg1$ and/or $C'\gg C$. In the regime
   $T<T_{\rm BKT}$, the system shows activation behavior with the gap
   $\Delta=T_{\rm BKT}\ln(\xi_{\rm s}^{\;2})$.}
\label{fig2}
\end{figure}

Using the results of the previous sections, we are now in a position
to discuss the phase diagram of the array.  An array having
inter--grain capacitances $C'$ only exhibits a BKT phase transition
between the low--temperature insulating and the high--temperature
conducting phases. Its phase diagram on the plane temperature vs. bare
inter--grain conductance, $g$, is shown in Fig.~\ref{fig2}. Unlike
previous works~\cite{fs2,fs} that predicted a zero--temperature metal
for $g>g_c\simeq 1$, we find that the low--temperature phase is an
insulator for arbitrarily large $g$. The critical temperature, $T_{\rm
  BKT}(g)$, however, drops sharply at $g\simeq 1$ and, at large $g\gg
1$, behaves as $T_{\rm BKT}\sim E_cg \exp\{-g\}$.  As shown in
Appendix~\ref{app-phase}, the disagreement is not a consequence of the
different model we use, but can be traced back to the disregard of the
quantum fluctuations of phase in the earlier works.  By contrast,
Ref.~[\onlinecite{tschersich}] uses a perturbative renormalization
scheme that neglects instanton configurations.  However, it is
precisely these instanton configurations that reflect the discreetness
of charge which is the key point in identifying the transition.

In the presence of the self--capacitance, $C>0$, the screening length
$\xi_{\rm s}\simeq \sqrt{1+C'/C}\, \exp\{g/2\}$ is finite, and the
transition is smeared into a crossover. The crossover is sharp as long
as $\xi_{\rm s}\gg 1$. Regardless of the ratio $C/C'$, this is the
case for $g\gg 1$.  In this regime the charge gap is parametrically
larger than the crossover temperature, $\Delta\simeq gT_{\rm BKT}$,
and therefore the residual conductivity below the crossover, though
finite, is exponentially small, $\sigma\lesssim g\exp\{-g\}$. As a
result, quantitatively, there is little difference between the models
with and without self--capacitance. This is not the case for
$g\lesssim 1$: unless $C'\gg C$, the crossover is gone, and the
conductivity follows the simple activation law $\sigma\simeq
g\exp\{-\Delta/T\}$ with $\Delta \simeq E_c$.

\section{Finite gate voltage}
\label{sec-gate}

So far we have restricted ourselves to the case of zero gate voltage
only. A finite gate voltage induces a continuous background charge
$q\propto V_{\rm gate}$ on the grains. In this case the charging term
in the action Eq.~(\ref{eq-classical}) has to be replaced with $S_{\rm
  cl}^{\rm (c)}[\vec\theta;q]=E_c/T \sum_{\bf
  l}(\nabla\cdot\vec\theta_{\bf l}-q)^2$. Alternatively one may shift
the $\vec\theta$ field by $ q{\bf l}$ to move the $q$--dependence into
the pinning term $\frac{\gamma(T)}{2\pi^2}\cos(2\pi(\theta_i+ql_i))$.
Since grain coordinates $l_i$ take only integer values, the model is
periodic in the $q$--space with unit period. In this work we restrict
ourselves to a uniform gate voltage $q({\bf l})=q=const$, leaving
considerations of a random background charge for future studies.

For small $q$, the system is in a particle--hole symmetric ``neutral''
state: the ground state is still (as for $q=0$) characterized by
$\vec\theta_{\bf l}=0$.  At some finite value $q=q^*$, a transition
takes place, where the ground state becomes charged (with a
non--integer average number of electrons per dot) and spatially
non--uniform.  To find $q^*$, let us compute the soliton energy in the
presence of $q$.  Since the $q$--dependence of the Hamiltonian is a
pure boundary effect, one immediately finds the soliton energy
$\Delta(q)=\Delta(0)-2qE_c$. At $q^*=\Delta(0)/(2E_c)$ the soliton
energy $\Delta(q)$ vanishes.  This marks the transition into the
charged state: for $q>q^*$, solitons are created at no cost.

In the charged phase, the density of solitons in the array is finite
even at zero temperature.  In order to find the soliton density at
$q>q^*$, one has to take into account interaction between the
solitons. At small densities, $n_{\rm s}<\xi_{\rm s}^{-2}=\gamma$, the
interaction between solitons is exponentially weak. The energy cost
associated with a soliton density $n_{\rm s}$ reads
\begin{eqnarray}
E_<(n_{\rm s})=n_{\rm
  s}\Delta(q)+\frac2\pi E_c  \gamma \,n_{\rm s}
K_0\left(\sqrt{\frac\gamma{n_{\rm s}}}\right),
\end{eqnarray}
where the second contribution is given by the interaction energy of a
pair of solitons separated by the distance $1/\sqrt n_{\rm s}$,
multiplied by the soliton density. Since $\sqrt{\gamma/n_{\rm
    s}}\gg1$, we can use the asymptotic expression for the Bessel
function, $K_0(x)\sim x^{-1/2}\exp\{-x\}$ for $x\to\infty$. The
optimal density is determined by the minimum of $E_<(n_{\rm s})$.
Minimization of $E_<(n_{\rm s})$ with respect to $n_{\rm s}$ yields
$n_{\rm s}(q)\sim\gamma/\ln^2[\gamma/(q-q^*)]$, where we used that
$\Delta(q)=2E_c(q^*-q)$.  Thus, at $q\gtrsim q^*$, the soliton density
rises rapidly until at $q-q^*\simeq\gamma$, the distance between
solitons reaches $\xi_{\rm s}$. For $n_{\rm s}>\xi_{\rm s}^{-2}$, the
solitons start to interact logarithmically.  Consequently, the
expression for the energy has to be modified as
\begin{eqnarray}
E_>(n_{\rm s})\eq n_{\rm
  s}\Delta(q)+\frac1{2\pi} E_c\gamma \,n_{\rm
  s}^2\!\!\int\limits_{1/\sqrt{n_{\rm s}}}^\infty \!\!l\,dl\;
K_0\left(\sqrt\gamma \,l\right)\nonumber\\
&\simeq& n_{\rm
  s}\Delta(q)+ \frac1{2\pi}E_c n_{\rm s} \left(\!n_{\rm s} - \gamma
  \,\ln\!\sqrt{\frac{n_{\rm s}}\gamma}\right)\!, 
\end{eqnarray}
where the second contribution describes the interaction energy of the
solitons with density $n_{\rm s}$; in the volume $\xi_{\rm
  s}^2=1/\gamma$, the interaction is logarithmic [$K_0(x)\simeq -\ln
x$ for $x\ll1$].  In this regime, the minimization yields $n_{\rm
  s}(q)\sim 2\pi(q-q^*)+ (\gamma/4)\ln[( q-q^*)/\gamma]$ for
$q-q^*\gg\gamma$.

Naively, one would expect the system to be no longer insulating once
the density of solitons becomes finite at $q>q^*$ -- which would be
the case if the solitons were mobile.  However, even though the
soliton density in the system is finite, $n_{\rm s}>0$, it turns out
that -- except for a narrow region $q-q^*<\gamma$, where the
interaction between solitons is exponentially weak -- the solitons
form a Wigner crystal which is pinned due to the underlying lattice
structure.  Thus, transport is still activated.

To understand this fact, we use the analogy with the formation of
vortices in a type II superconducting film~\cite{deGennes}. The field
$\vec\theta$ may be viewed as ${\bf A}\times{\bf n}_z$, where ${\bf
  A}$ is the vector potential and ${\bf n}_z$ is a unit vector normal
to the film. Since the local magnetic field is given as ${\bf
  h}=h\,{\bf n}_z=\nabla\times{\bf A}$, the correspondence goes as
$\nabla\!\cdot\!\theta=h$ and the charge quantization in the array is
equivalent of the flux quantization in the superconductor,
$\int(d^2l)\,h=k$ ($k\in\Bbb{Z}$), where $h$ is measured in units of
the flux quantum $\phi_0$.  In this analogy, the gate voltage
translates to the external magnetic field, $H$, and the gate voltage
$q^*$ corresponds to the critical magnetic field $H_{c1}$, where it
becomes energetically favorable to create vortices.  The
correspondences are summarized in the following ``dictionary'':\\[-0.4cm]
\begin{center}
\begin{tabular}{cc|cc}
array &&& superconducting film\\[0.1cm]
\hline
&&&\\[-0.3cm]
$\vec\theta$ &&& ${\bf A}\times{\bf n}_z
= -{\lambda^2}\nabla h$\\
charge $\nabla\cdot\vec\theta$ &&&  local magnetic
field $h$\\
$\xi_{\rm s}=1/\sqrt{\gamma}$ &&&  penetration depth $\lambda$ \\
background charge $q$ &&& external magnetic field $H$\\
$q^*$ &&& $H_{c1}$\\[-0.1cm]
\end{tabular}
\end{center}
Above $H_{c1}$ there is a finite density of vortices in the system
which at low enough temperatures form an Abrikosov
lattice~\cite{abrikosov}. In a clean film, the vortex lattice is free
to move, but it is easily pinned by the system boundaries, the
underlying lattice structure (as in Josephson junction arrays) or any
sort of disorder.  Upon increasing the temperature the vortex lattice
eventually melts~\cite{BKT-kt,Tm}, and above the melting temperature
$T_{\rm m}$ most of the vortices are free to move.  The melting
temperature at finite $H>H_{c1}$ is smaller than, but parametrically
the same as the Berezinskii--Kosterlitz--Thouless temperature at zero
magnetic field~\cite{Tm}. Thus, at $T<T_{\rm m}$ the system is
superconducting while at $T>T_{\rm m}$ the moving vortices lead to
dissipation.

 \begin{figure}[h]
   \centerline{\epsfxsize=3.25in\epsfbox{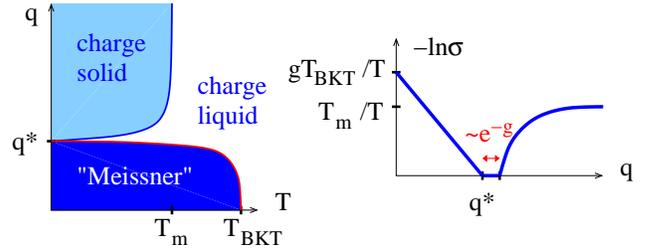}}
 \caption{System properties as a function of an external gate
   voltage. The system becomes charged at $q>q^*$. If the soliton
   density is larger than $1/\xi_{\rm s}^{\;2}$, charges arrange into
   a (pinned) Wigner crystal. a) Phase diagram. b) Conductivity at
   $T<T_{\rm m}$ as a function of $q$.}
 \label{fig3}
 \end{figure}
 
 Translating back to our problem this means that at $q>q^*$ the
 solitons form a Wigner crystal once their density is sufficiently
 large such that the interaction is logarithmic. Only in the narrow
 interval $q^*< q< q^* + \xi_{\rm s}^{-2}$ the system is in the
 conducting charge liquid state. Upon increasing the gate voltage, the
 Wigner crystal forms and, due to lattice pinning, the system is an
 insulator at temperatures smaller than the melting temperature.  The
 latter is of the order of $T_{\rm BKT}$. Note that while for $q<q^*$
 charge is carried by individual (thermally-activated) solitons, for
 $q>q^*$ the mobile charges are lattice defects, whose core energy is
 proportional to the logarithm of the lattice constant of the Wigner
 crystal.

\section{Conclusions}
\label{sec-conclusions}

The conductivity of a granular material with small ($g\ll 1$)
inter--grain conductances is controlled by the Coulomb blockade effect
in separate grains~\cite{fs}. Charge transport in such an array occurs
by electron hops between single grains. Because of the Coulomb
blockade, the granular array behaves as an insulator at low
temperatures. The characteristic energy determining the activation of
the charge transport, is associated with the single--grain charging
energy $E_c$.

In this paper, we investigated the properties of a granular array at
high inter--grain conductance, $g\gg 1$. We concentrated on the
simplest model, neglecting the spacing between the discrete electron
levels in the grains ($\delta\to 0$), and in the main part of the
paper we also assumed the ideal limit of zero background charge
($q=0$).  We found that the granular array at sufficiently low
temperatures remains an insulator even in the case of $g\gg 1$. The
large inter--grain conductance, however, does affect the nature of the
charge carrier.  Instead of an integer (in the units of $e$)
uncompensated charge sitting on a single grain, it is rather a
charge--$e$ soliton involving many grains.  There is a sharp crossover
to low conductance at temperature the $T_{\rm BKT}\propto
E_c\exp\{-g\}$. Below the crossover, the electron transport is
associated with the activation of solitons with charges $\pm e$; the
corresponding activation energy is relatively high, $\Delta\simeq
gT_{\rm BKT}$. The approach to the crossover region from the
high--temperature side can be described in terms of the correlation
radius for the Berezinskii--Kosterlitz--Thouless transition, see
Eq.~(\ref{eq-above}).  Comparison of Eqs.~(\ref{eq-res}) and
(\ref{eq-above}) shows that the crossover width is $\delta T\sim
T_{\rm BKT}/g^2\ll T_{\rm BKT}$.

The effect of a gate voltage $q$ applied uniformly throughout the
array is to much extent similar to that of an external magnetic field
applied to a type II superconductor. Until $q$ reaches a certain
critical value $q^*$ the system's behavior does not change
qualitatively (that is, it exhibits the BKT crossover from insulator
to metal). The charge gap, $\Delta$, and the crossover temperature,
$T_{\rm BKT}$, decrease with increasing $q$. At $q=q^*$ the gap
vanishes (while the crossover temperature remains finite), and for
even larger gate voltages there is a certain ground--state density,
$n_{\rm s}=n_{\rm s}(q)$, of charge solitons. As long as this density
is small $n_{\rm s}(q)<\xi_{\rm s}^{-2}$ the solitons are in a liquid
state and the array conducts.  At larger density, $n_{\rm
  s}(q)>\xi_{\rm s}^{-2}$, the solitons form a Wigner crystal pinned
by the lattice.  As a result, the array is again in the insulating
state with the charge gap determined by the cost of a defect in the
Wigner crystal.

Finally, let us mention related issues that are {\em not} addressed in
the present paper. The first one is the role of disorder.  The most
relevant is charge disorder equivalent to a grain--dependent gate
voltage $q_{\bf l}$. In the extreme scenario one may assume that
$q_{\bf l}$ are independent random variables, uniformly distributed in
$[0,1]$. One would then like to solve the classical statistical
problem formulated by Eqs.~(\ref{eq-langevin}) and (\ref{eq-noise})
with random $q_{\rm l}$ in the argument of $\sin(2\pi (\theta_{i,{\bf
    l}} + q_{\bf l}))$. Despite many similarities with the vortex
physics, one can not simply transfer the known results from the pinned
vortex lattice literature~\cite{blatter}. The reason is that random
charge, $q_{\bf l}$, translates into a strong (of order of $H_{c2}$)
fluctuating magnetic field, rather a than fluctuating pinning
potential.

Another unaddressed issue is the role of quantum coherence, which
enters the problem through the mean level spacing $\delta$. Our
results are valid as long as $g^2\delta\lesssim T_{\rm BKT}$. In the
opposite limit, the quantum coherence effects (most notably Anderson
localization) start to interfere with the effects of
electron--electron interactions, considered here. One may expect that
both effects drive the system towards the insulating ground--state.
(It is worth mentioning that in both cases the characteristic length
happens to be exponentially large in the bare conductance, $g$.)  The
structure of such insulator is not known currently.

\begin{acknowledgments}
  We are grateful to A. Altland for valuable discussions.  This work
  was supported by NSF grants DMR01-20702, DMR02-37296, and
  EIA02-10736. Furthermore, JSM was supported by a Feodor Lynen
  fellowship of the Humboldt Foundation.  AK and JSM acknowledge the
  hospitality of KITP (UCSB), where part of this research was done
  under NSF grant PHY99-07949.
\end{acknowledgments}

\begin{appendix}

\section{Multi--channel contacts}
\label{app-gamma}

In the main text (Sec.~\ref{subsec-array}), we derived a classical
model for the single--channel case. Here we discuss its generalization
to $N\geq2$ channels.  For every channel $\alpha=1,\ldots,N$ of a
multi--channel contact, one introduces a field $\theta_\alpha(\tau)$.
Consider an $M\times M$ array with $N$ channels in each of the $2M^2$
contacts. The quadratic part of the action reads
\begin{eqnarray}
S_2 = \frac1T\sum_{{\bf l},m}\left(\pi|\omega_m|\sum_\alpha
    \vec\theta_{{\bf l},\alpha}^{\;2}+E_c\Big(\sum_\alpha
    \nabla\cdot\vec\theta_{{\bf l},\alpha}\Big)^2\right)\!,
\end{eqnarray}
while the backscattering is described by
\begin{eqnarray}
S_r = -\frac{E_c}\pi\sum_{{\bf l},\alpha} \sum_{i=x,y}r_{i{\bf
    l}\alpha}\! \int\limits_0^\beta\! d\tau\,
  \cos(2\pi\theta_{i,{\bf l},\alpha}),
\end{eqnarray}
where the high--energy modes $E_c<|\omega_m|<D$ have already been
integrated out. [At energies larger than $E_c$, all modes are
decoupled and, thus, can be integrated out for each channel
separately.] Here, in order to clarify the following evaluation
scheme, the reflection coefficients $r$ have been given indices
specifying the direction, contact, and channel.

Only the $2M^2$ symmetric modes $\theta_{\bf
  l}=\sum_\alpha\theta_{{\bf l},\alpha}$ couple to external
parameters, such as gate voltages.  We want, thus, to find an
effective action for $\theta_{\bf l}$ by integrating out $2M^2(N-1)$
asymmetric modes. To this end let us change variables from
$\theta_{{\bf l},\alpha}$ ($\alpha=1\dots N$) to $\theta_{\bf l}$ and
$\tilde \theta_{{\bf l},\alpha}=\theta_{{\bf l},\alpha}-(\theta_{\bf
  l}-\sum_{\alpha'>\alpha}\theta_{{\bf l},\alpha'})/(\alpha+1)$
($\alpha,\alpha'=1\dots N-1$). While the symmetric fields $\theta_{\bf
  l}$ are massive due to the charging term, all the asymmetric fields
$\tilde \theta_{{\bf l},\alpha}$ are massless. As a result, the
perturbation theory in powers of $r_{i{\bf l}\alpha}$ contains only
the terms that do not have massless fields $\tilde \theta_{{\bf
    l},\alpha}$ in the exponents (cosines).  Rewriting the
backscattering action in terms of the new fields, one can see that the
lowest order non--vanishing terms are of the order
$\prod_{\alpha=1}^Nr_{i{\bf l}\alpha}$, where the product runs over
{\em all} channels of a given contact:
\begin{widetext}
\begin{eqnarray}
Z_N\sim E_c^N\prod_{\alpha=1}^Nr_{i{\bf l}\alpha}\int
d\tau_\alpha\,\cos\left(\frac{2\pi}N\sum_\alpha \theta_{i,{\bf
      l}}(\tau_\alpha)\right) \prod_{\alpha=1}^N \left\langle
  \exp\left\{2\pi 
  i\left(\tilde\theta_{i,{\bf l}\alpha}(\tau_\alpha) - 
    \frac1\alpha\left(\tilde\theta_{i,{\bf l},\alpha}(\tau_N) +
      \sum_{\alpha'<\alpha}\tilde\theta_{i,{\bf
          l},\alpha}(\tau_{\alpha'}) \right)\right)\right\}\right
\rangle_{\tilde\theta_{{\bf l},\alpha}}\!\!\!.\nonumber
\end{eqnarray}
\end{widetext}
Taking the averages $\langle\dots\rangle_{\tilde\theta_{{\bf
      l},\alpha}}$ with the actions
\begin{eqnarray}
S[\theta_{i,{\bf
    l},\alpha}] = \frac1T \sum_m \frac{\alpha+1}\alpha\pi
|\omega_m|\tilde\theta_{i,{\bf
    l},\alpha}^2\, ,
\end{eqnarray}
yields $E_c^{1-N}\prod_{\alpha=1}^N \prod_{\alpha'>\alpha}
(\tau_\alpha-\tau_{\alpha'})^{-2/N}$ for the product of correlators
$\prod_\alpha\langle\dots\rangle_{\tilde\theta_{{\bf l},\alpha}}$.

Thus, the effective action for $\vec\theta_{\bf l}$ reads
\begin{widetext}
\begin{eqnarray}
S[\vec\theta]\!=\!\sum_{\bf l}\left\{\!\frac1T\sum_m\left( \frac\pi N
    |\omega_m| \vec\theta_{{\bf l}}^{\;2}+E_c\Big(
    \nabla\!\cdot\!\vec\theta_{\bf l}\Big)^2\right) -
  \frac{E_c}\pi\! \sum_{i=x,y}
  \prod_{\alpha=1}^N r_{i{\bf l}\alpha}\!\! \int\! d\tau_\alpha
  \prod_{\alpha'>\alpha}
  \frac1{(\tau_\alpha\!-\!\tau_{\alpha'})^{2/N}}
  \cos\Big(\frac{2\pi}N \sum_\alpha\theta_{i,{\bf
      l}}(\tau_\alpha)\Big) \!\right\}.
\end{eqnarray}
\end{widetext}
Note that it is important to keep the non--local in time structure of
the cosine--term.

At this stage, we can proceed to integrate out all the remaining modes
except the static one, $\theta_{m=0}$ -- as in the single--channel
case. The prefactor of the cosine--term $V_0=({E_c}/\pi) \prod_\alpha
r_{i{\bf l}\alpha}$ is renormalized according to
\begin{eqnarray}
V_0&\to&V(T)=V_0\,\exp\!\Big\{-\frac{2\pi^2}{N^2}
\sum_{\alpha,\alpha'} \langle
\theta(\tau_\alpha)\theta(\tau_{\alpha'})
\rangle_{\theta_{m\neq0}}\Big\}\nonumber\\
&=& V_0\,\exp\!\Big\{-\!\sum_{m\neq0} f(\omega_m)\big(1 \!+\!
\frac2N\sum_{\alpha,\alpha'>\alpha}
\cos\omega_m\tau_{\alpha\alpha'}\big)\Big\},\nonumber
\end{eqnarray}
where $f(\omega_m)=T/(4M^2)\sum_{\bf q}\{(E_{\bf
  q}+\pi|\omega_m|)^{-1}+(\pi|\omega_m|)^{-1}\}$; see
section~\ref{sec-charge}. Since typical time differences
$\tau_{\alpha\alpha'}=\tau_\alpha-\tau_{\alpha'}$ are of the order
$1/T$ (the time integrals are dominated by the upper limit of
integration), the last cosine--term inside the exponent may be
disregarded.  As a result we find $V(T)=V_0\sqrt{T/E_c}$.

Finally, one may perform the multiple time integrations in the
prefactor of the cosine. The integral over the center--of--mass time
$\tau=\sum_\alpha\tau_\alpha/N$ contributes a factor $1/T$, while the
integration over $N-1$ independent time differences
$\tau_{\alpha}-\tau$ yields a constant $c_N$ multiplied by the
logarithmic factor~\cite{ABG} $\ln E_c/T$. The latter follows simply
from power counting. The same logarithmic factor appears in the
framework of the phase model, Appendix \ref{app-phase}, as a result of
zero--mode integration. Since all our evaluations of $\gamma$ are done
up to a numerical factor, we shall not keep this logarithm explicitly.
We, thus, reproduce Eq.~(\ref{eq-classical}) with $\gamma(T)\sim
\sqrt{T/E_c}\, \prod_{\alpha=1}^N r_\alpha$. Continuation to $T<T_0$
follows the same way as discussed in the main text for the
single--channel case.

\section{Phase model}
\label{app-phase}

In this Appendix, we establish correspondence between the charge
representation, employed in the paper, and the more commonly used
phase model. The latter may be straightforwardly derived starting from
the fermionic tunneling Hamiltonian. Integration over the fermionic
degrees of freedom (under the assumption of vanishing level spacing in
every grain, $\delta\to 0$), leads to a model formulated in terms of
the dynamic phase variable~\cite{AES}, $\phi_{\bf l}(\tau)$. Its time
derivative, $\dot\phi_{\bf l}(\tau)$, has the meaning of a fluctuating
instantaneous voltage on grain ${\bf l}$. The resulting action is a
straightforward generalization of the Ambegaokar-Eckern-Sch\"on (AES)
action~\cite{AES} to the array geometry.  It consists of the charging
term,
\begin{equation}
                                           \label{AEScharging}
S_{\rm c}[\phi]=\int\limits_0^{\beta}\!\! d\tau \sum_{\bf
  l}\left[ {\dot\phi_{\bf l}^{\;2}\over 4E_c} -  i q
  \dot\phi_{\bf l}\right]\, ,
\end{equation}
and the dissipative term
\begin{equation}
\label{eq-AES}
S_{\rm d}[\phi]=\frac {g
      T^2}4\int\!\!\!\int\limits_0^{\beta}\!\! d\tau  d\tau'
    \sum_{\langle{\bf l},{\bf l'}\rangle}
  \frac{\sin^2(\phi_{\bf ll'}(\tau)-\phi_{\bf ll'}(\tau'))}{\sin^2(\pi
  T(\tau-\tau'))}\, ,
\end{equation}
describing tunneling between nearest neighbor grains $\langle{\bf
  l},{\bf l'}\rangle$. Here, $\phi_{\bf ll'}=(\phi_{\bf l}-\phi_{\bf
  l'})/2$.

\begin{figure}[h]
  \centerline{\epsfxsize=2.5in\epsfbox{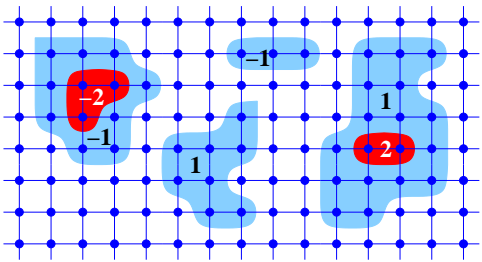}}
 \caption{A typical island configuration is shown. The
   numbers correspond to winding numbers in the phase model.}
\label{fig4}
\end{figure}

The phase field $\phi_{\bf l}(\tau)$ obeys the boundary condition
$\phi_{\bf l}(\beta)-\phi_{\bf l}(0)=2\pi W_{\bf l}$, where $W_{\bf
  l}\in \Bbb{Z}$ is an integer called winding number. In addition to
the trivial configuration $\phi_{\bf l}=0$, the stationary
configurations of the dissipative action, $S_{\rm d}$, are given by
Korshunov instantons~\cite{korshunov}
\begin{equation}
e^{i\phi_{\bf l}^{(W_{\bf l})}}=\prod_{a=1}^{|W_{\bf
l}|}\frac{e^{2\pi i\tau
    T}-z_{a}}{1-\bar z_{a}e^{2\pi i\tau T}}
\end{equation}
characterized by the spatially--dependent winding number $W_{\bf l}$
and a set of complex parameters $|z_{a}|<1$. In the regime $T\ll E_c$,
these configurations are a good approximation to the saddle points of
the total action. In the same approximation the $z_{a}$ are zero--mode
coordinates: the instanton action is almost $z$-independent (safe for
the charging terms that weakly depends on $z$).  Neglecting this
dependence, the action for a certain winding number configuration,
$\{W_{\bf l}\}$, reads
\begin{eqnarray}
\label{eq-sw}
S_W \!\simeq\! \frac{\pi^2T}{E_c}\sum_{\bf
  l} W_{\bf l}^2
+\frac g4\sum_{\langle{\bf l},{\bf l'}\rangle}|W_{\bf
  ll'}|,
\end{eqnarray}
where $W_{\bf ll'}\equiv W_{\bf l}-W_{\bf l'}$. In the regime, we are
interested in, $T\ll E_c$ and $g\gg 1$, the dominant contribution
comes from the second (tunneling) term in this expression.  Since the
latter depends only on the differences of winding numbers on
neighboring grains, it favors configurations with spatially extended
regions with a fixed constant winding number, e.g. $W_{\bf l}=\pm 1$
for a closed set of grains ${\bf l}$. We shall refer to such sets of
grains with a fixed non--zero winding number as ``islands''. A typical
phase--field configuration contains, therefore, a number of
``islands'' (with fixed non--zero windings) embedded in the sea of
$W=0$ grains. An example of such a configuration is shown in
Fig.~\ref{fig4}. According to Eq.~(\ref{eq-sw}), the action cost of
one such island with the winding number $W$ is $S= A W^2(\pi^2 T/E_c)
+ L|W|(g/2)$, where $A$ is the area of the island (number of grains
inside) and $L$ is its circumference (number of contacts with a
winding number jump across them).

The same picture was employed in previous studies of 2d granular
arrays~\cite{fs2,fs}, where the island structure was mapped onto the,
so called, solid--on--solid model. The latter is known to exhibit a
BKT transition. [A true transition takes place if there is no cost for
the island's area, as is the case for the model with mutual
capacitances only. Indeed, in such a model both charging and tunneling
terms provide a cost proportional to the island's circumference, $L$.
In presence of the self--capacitance (and thus an area--proportional
cost) the transition is smeared into a crossover.] What was missed in
the previous studies is an account of fluctuations on top of the
stationary island--like configurations.

We provide such an account here. Consider a stationary configuration
consisting of a single island with a fixed winding number $W$.
Expanding to the second order in deviations $\phi_{\bf l}=\phi_{\bf
  l}^{(W_{\bf l})}+\varphi_{\bf l}$, one finds for the fluctuating
part of the action
\begin{equation}
                                      \label{eq-fluctuations}
\delta S = \sum\limits_m \sum\limits_{\bf l,l'} \bar\varphi_{{\bf
l},m} M_{{\bf l,l'},m}^{(W)}\, \varphi_{{\bf l'},m}\, ,
\end{equation}
where $m$ is a Matsubara index, and $M_{{\bf l,l'},m}^{(W)}\equiv{\bf
  M}_m^{(W)}$ is the fluctuation matrix. The fluctuation factor
associated with this configuration is given by
\begin{equation}
                                        \label{eq-determinant}
\prod\limits_{m>1}\frac{\det {\bf M}_m^{(0)}}{\det {\bf
M}_m^{(W)}}=\exp\left\{-\sum\limits_m \mbox{Tr}\ln \left({\bf
M}_m^{(0)}\right)^{-1}{\bf M}_m^{(W)}   \right\},
\end{equation}
where ${\bf M}_m^{(0)}$ is the corresponding fluctuation matrix for
the flat ($W=0$) stationary configuration. In the regime $g\gg 1$, the
dominant fluctuation contribution comes from the expansion of the
tunneling term, $S_{\rm d}$. This leads to $M_{{\bf l,l'},m}^{(0)}=
-g|\omega_m|$ for nearest neighbors $\langle {\bf l,l'}\rangle $,
while $M_{{\bf l,l},m}^{(0)}=-\sum_{{\bf l'}\neq {\bf l}} M_{{\bf
    l,l'},m}^{(0)}$ and $M_{{\bf l,l'},m}^{(0)}=0$ otherwise. In
presence of the island, the off-diagonal elements of the fluctuation
matrix are changed to $M_{{\bf l,l'},m}^{(W)}=-g|\omega_{m-|W|}|$ (and
the diagonal elements accordingly), only if ${\bf l}$ and ${\bf l'}$
are nearest neighbors laying across the island's boundary. As a
result, one may write ${\bf M}_{m}^{(W)}={\bf M}_{m}^{(0)} - |W| {\bf
  \delta M}$, where the matrix ${\bf \delta M}$ has entries $\pm 2\pi
Tg$ along the island's boundary and zeros everywhere else.  Returning
to the calculation of the fluctuation factor,
Eq.~(\ref{eq-determinant}), one finds $\mbox{Tr}\ln \left({\bf
    M}_m^{(0)}\right)^{-1}{\bf M}_m^{(W)} = \mbox{Tr}\ln\!\left[1-
  |W|\left({\bf M}_m^{(0)}\right)^{-1}\!\!\!{\bf \delta M}\!\right]
\!\approx\! -|W|\mbox{Tr}\left({\bf M}_m^{(0)}\right)^{-1}\!\!\!{\bf
  \delta M}$. Higher order terms in the expansion of the logarithm are
rapidly convergent upon Matsubara summation and, therefore, may be
safely neglected. Since ${\bf M}_m^{(0)} \sim g|\omega_m|$, summation
over the Matsubara index in $\sum_m |W|\mbox{Tr}\left({\bf
    M}_m^{(0)}\right)^{-1}\!\!{\bf \delta M}$ leads to the logarithmic
divergence \cite{Grabert96}. It is cut off by the charging part of the
action at $m\approx gE_c/T\gg 1$. The summation (trace) over spatial
indices results in a factor proportional to $L$, the island's
circumference, as it counts the number of non--zero entries in ${\bf
  \delta M}$. Finally, a careful evaluation of the numerical
coefficient~\cite{unpub} leads to the fluctuation factor,
Eq.~(\ref{eq-determinant}), equal to $(gE_c/T)^{L|W|/2}$.

As a result, an island of winding number $W$ with area $A$ and
circumference $L$ contributes to the partition function of the model
with the relative factor
\begin{equation}
                      \label{eq-factor}
P_W(A,L)\!=\! \left(e^{-\pi^2T/E_c}\right)^{A W^2}
\!\!\left(\sqrt{gE_c\over T}\, e^{-g/2} \right)^{L|W|}\!.
\end{equation}
(Actually, the statistical weight of an island contains also a factor
$(\ln E_c/T)^{|W|}$, coming from the zero--mode, $z_a$, integrations
\cite{Grabert96}. This factor has its exact analog in the charge
model, mentioned at the end of Appendix \ref{app-gamma}. Hereafter we
omit it for brevity.)

We shall show now that the perturbative expansion in powers of
$\gamma(T)$ of the classical model, Eq.~(\ref{eq-classical}), leads to
the same island picture. In this case, every island carries the
relative factor $\tilde P_W(A,L)\!=\!  \left(e^{-\pi^2T/E_c}\right)^{A
  W^2} \!\left(E_c\gamma(T)/ (2\pi^2 T)\right)^{L |W|}$. We can, thus,
identify the two models provided $\gamma(T) \simeq \sqrt{gT/E_c}\,
e^{-g/2}$. Notice that $\gamma(T)\propto \sqrt{T/E_c}$ is exactly what
one expects for the high--temperature, $T>T_0$, charge model. At lower
temperature, non--linear fluctuation corrections in the phase model
diverge~\cite{tschersich} and the above treatment runs out of
validity. However, having establish the equivalence of the phase and
charge models at $T>T_0$, one may proceed with the analysis of the
latter even at smaller temperatures.

To complete the proof, we elucidate now the island structure of the
perturbative expansion of the charge model, Eq.~(\ref{eq-classical}).
Consider the expansion of the partition function ${\cal Z}=\int
D\vec\theta\;\exp\left\{-S[\vec\theta]\right\}$, with the action
$S[\vec\theta]$ given by Eq.~(\ref{eq-classical}), in powers of the
small parameter $\gamma$. The partition function can be written as
${\cal Z}=\sum_{n=0}^\infty Z_n (E_c\gamma/T)^{n}$, where $Z_n$ is a
product of $n$ cosine terms averaged with the action $S_c[\vec \theta]
=E_c\sum_{\bf l}(\nabla\cdot\vec \theta_{\bf l})^2/T$.  There are two
types of contributions: a) terms with higher powers of the cosine
taken at the same link and b) terms involving different links. The
first class of terms describes perturbative corrections to the
conductance of a single contact, and may be shown to be equivalent to
those of Ref.~[\onlinecite{tschersich}] in the framework of the phase
model. The second class corresponds to the instanton terms and is the
subject of our focus. These terms exhibit ``island formation''. To
illustrate this, let us label the coefficients $\gamma_{i,{\bf l}}$
(even though we assume them all to be equal), where $i=x,y$. Terms of
the form $\prod_{\underline{\alpha}}\gamma_{\underline{\alpha}}$ are
non--zero only if the lines crossing all contacts
${\underline{\alpha}}=(i,{\bf l})$ form closed loops, see
Fig.~\ref{fig4}. I.e. the lowest-order non-local term is proportional
to $\gamma^4=\gamma_{x,{\bf l}}\gamma_{x,{\bf l+e}_x}\gamma_{y,{\bf
    l}}\gamma_{y,{\bf l+e}_y}$ -- involving all the four links
surrounding grain ${\bf l}$.  This property of the model is due to the
presence of massless modes, as is explained below.

Rewriting $\vec\theta=\nabla{\chi}+\nabla\times\eta$, one finds that
the charging action takes the form $S_c[\chi,\eta] =E_c\sum_{\bf
  l}(\nabla \chi_{\bf l})^2/T$ and is thus $\eta$--independent. The
rotational field $\eta$ is therefore strictly massless. As a result,
as long as an argument of the cosine (exponential) function contains
the $\eta$--field, it averages to zero. Indeed, to obtain $Z_n$ one
has to average expressions of the form $\exp\{2\pi
i(\sum_{j_x=1}^s(\pm)\theta_{x,{\bf
    l}_{j_x}}+\sum_{j_y=s+1}^n(\pm)\theta_{y,{\bf l}_{j_y}}\}$, where
$j_i$ labels contacts in $i$--direction. The terms containing $\eta$
in the argument of this exponent, vanish. Therefore non--vanishing are
only those terms that have $\sum_{j_x}\pm(\eta_{\,{\bf
    l}_{j_x}}-\eta_{\,{\bf l}_{j_x}\!\!-{\bf e}_y})
+\sum_{j_y}\pm(\eta_{\,{\bf l}_{j_y}}-\eta_{\,{\bf l}_{j_y}\!\!-{\bf
    e}_x})=0$. It can be seen that this condition corresponds to the
island structure. As a result, every island brings a factor
$(E_c\gamma/T)^L$, where $L$ is its circumference, that simply
reflects the order of the perturbation theory needed to create the
island. For a proper (i.e. island--like) term of the perturbation
theory, the averaging over the massive $\chi$--fields results in the
factor $\exp\{-\pi^2TA/E_c\}$, where $A$ is the area. Finally, the
integer index $|W|$ corresponds to the possibility of having a
non--zero term of the perturbation theory, where links surrounding an
island are included $|W|$ times each. We have shown, thus, that the
perturbation theory in the charge model, Eq.~(\ref{eq-classical}),
produces the same island structure as the instanton expansion of the
phase model -- with the same relative factors, Eq.~(\ref{eq-factor}).
This completes the proof of the equivalence of the two models and
provides the value of $\gamma(T)$, Eq.~(\ref{eq-gamma}), for $g\gg 1$.

\end{appendix}

\end{document}